\documentclass[RNAAS]{aastex62}
\usepackage{comment}

\begin{document}

\title{Embedding Reliability for Unsupervised Classification of Gamma Ray Burst progenitors from Prompt Gamma-ray Emission}

\correspondingauthor{Nicol\'o Cibrario}
\email{nicolo.cibrario@unito.it}

\author[0000-0003-3842-4493]{Nicol\'o Cibrario}
\affiliation{Istituto Nazionale di Fisica Nucleare, Sezione di Torino, Via Pietro Giuria 1, 10125 Torino, Italy}%
\affiliation{Dipartimento di Fisica, Università degli Studi di Torino, Via Pietro Giuria 1, 10125 Torino, Italy}

\author[0000-0002-6548-5622]{Michela Negro}
\email{michelanegro@lsu.edu}
\affiliation{Department of Physics \& Astronomy, Louisiana State University, Baton Rouge, LA 70803, USA}


\section{} 

\cite{Negro2025} introduced an innovative approach to classifying gamma-ray bursts (GRBs) employing a deep learning (DL) algorithm. The analysis exploits a new type of input for GRBs, the \textit{waterfall plots}, and a tailored self-supervised and DL algorithm, based on a \textit{convolutional autoencoder} architecture \citep{convae}.

Each GRB in the dataset is characterized by a set of 12 images, each composed of thousands of pixels, which contain GRBs core prompt emission information, such as duration, temporal variation, pulse structure, spectral hardness and evolution, and how these parameters relate. A set of autoencoders  compresses the information of each GRB into a compact latent representation which is ultimately combined into 30 numerical features \citep[see][for details]{Negro2025}. This latent space provides a lower-dimensional yet information-rich description of the GRB prompt emission. To visualize the overall structure of the dataset and explore potential clustering, a UMAP algorithm is further applied \citep{UMAP}, reducing the dimensionality from 30 down to two dimensions.

While the autoencoder allows a quantitative evaluation of its performance through the reconstruction error, the subsequent UMAP-based dimensionality reduction presents a more challenging interpretability. Indeed, UMAP is widely used for visualizing high-dimensional data, but the reliability of individual point placements in the low-dimensionality embedding space is often difficult to assess \citep{UMAP-1, UMAP-2}. While in \cite{Negro2025} this was done qualitatively by leveraging the location within the embedding of known information and properties of some GRBs, here we present a more quantitative method based on statistics.

We propose employing the recently developed statistical method scDEED \citep{scdeed} to assess the trustworthiness of each GRB position in the resulting 2D UMAP embedding. Here we report the step-by-step general functioning of the algorithm, and its application to the 30-dimensional dataset achieved by the convolutional autoencoder described in \cite{Negro2025}.

\begin{enumerate}
    \item \textbf{Dimensionality reduction with UMAP}  
    The procedure begins by applying a dimensionality reduction algorithm to the original high-dimensional dataset. In our case, we applied UMAP on the 30-dimensional GRB latent space, producing a two-dimensional embedding that facilitates visualization and further analysis.

    \item \textbf{Define pre- and post-embedding neighbors}
    For each data point, we identify its $k$ closest neighbors (with $k$ typically set to a fraction of the dataset size, e.g., 50\%). This neighborhood is computed twice: once in the original 30-dimensional latent space, and once in the two-dimensional embedding, using pairwise euclidean distances.

    \item \textbf{Evaluate Pearson correlation of distance vectors}
    For each point, two ordered vectors of distances are evaluated: one containing the distances in the final 2D space to the $k$ nearest neighbors in the original space, and one containing the distances in the final 2D space to the $k$ nearest neighbors in the same embedded space. The Pearson correlation between these two vectors defines a reliability score for that point, quantifying how well its neighborhood is preserved after embedding.

    \item \textbf{Generate a null distribution through permutation} To assess the significance of the reliability scores, a null model is constructed by randomly and independently permuting the 30 features of each GRB of the dataset, thereby destroying any meaningful neighborhood structure. Steps 1--3 are then applied to this permuted data, and the process is iterated (e.g., over 100 times), generating a null distribution of reliability scores that represents the behavior expected under random embeddings.

    \item \textbf{Flag trustworthy and dubious events} Reliability scores from the original dataset are compared against the null distribution. Events with scores above the 95th percentile are flagged as \emph{trustworthy}, while those below the 5th percentile are flagged as \emph{dubious}. Points falling between these thresholds remain unlabeled, reflecting an intermediate level of reliability.

\end{enumerate}

\begin{figure}[h!]
\begin{center}
\includegraphics[width=0.49\textwidth]{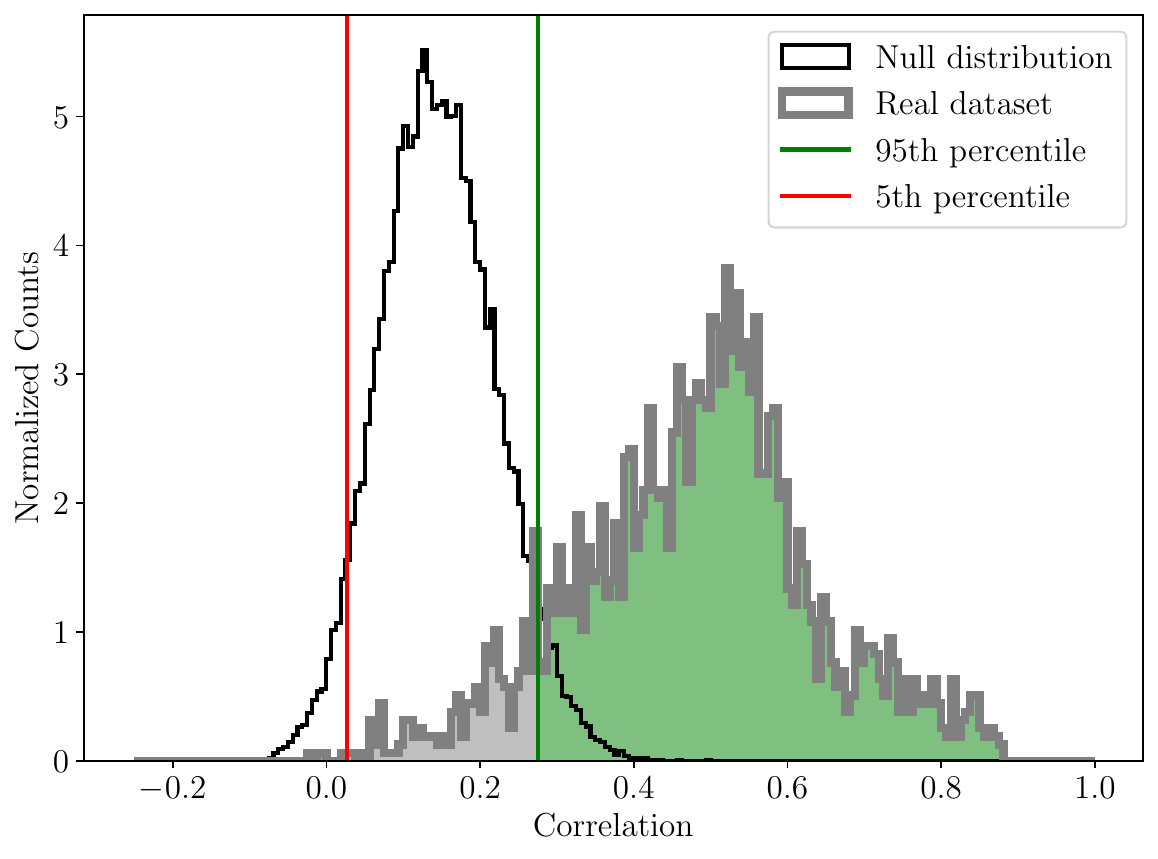}
\includegraphics[width=0.49\textwidth]{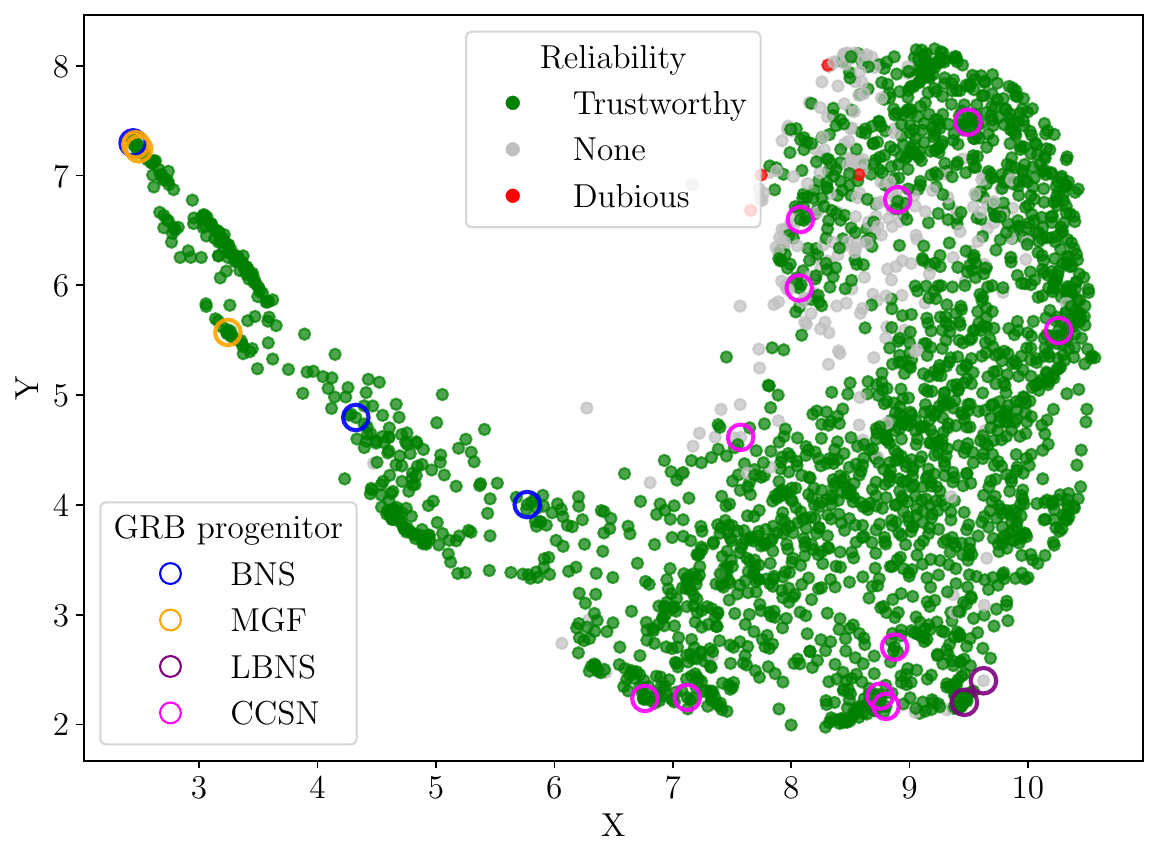}
\caption{Left panel: Pearson correlation values for each event, computed as in Step 3, are shown for the real dataset (gray histogram) and for the permuted datasets (black histogram), which serve as the null distribution. The 95th and 5th percentile thresholds are indicated by green and red lines, respectively. Right panel: Final 2D embedded distribution of the GRB dataset. Marker colors indicate the reliability measure: green for trustworthy, gray for intermediate, and red for dubious. Circled points indicate the GRB progenitor type: BNS $\rightarrow$ Binary neutron stars merger (blue), MGF $\rightarrow$ Magnetar giant flare (yellow), LBNS $\rightarrow$ Long binary neutron stars merger (purple), CCSN $\rightarrow$ Core-collapse supernova (pink).}
\label{figure}
\end{center}
\end{figure}

By combining local neighborhood preservation with a statistically principled null model, this method offers both an objective measure of trustworthiness for individual data points and a tool for optimizing embedding parameters. In doing so, it mitigates the risk of over-interpreting misleading structures in low-dimensional visualizations. The results of this method applied to our embedding are reported in Fig. \ref{figure}. The left panel displays the reliability scores of the GRB embedded distribution (gray thick line) with respect to the null permuted distribution (black line). The 95th and 5th percentile thresholds are reported respectively as green and red lines. The right panel reports the 2D final distribution, with each event colored with the respective reliability flag (green $\rightarrow$ trustworthy; gray $\rightarrow$ intermediate; red $\rightarrow$ dubious).

The results highlight the reliability of the event localization in the final 2D space: $\sim$90\% of the events are classified as trustworthy. None of the benchmark GRBs which were employed in \cite{Negro2025} to analyze the features of the final distribution are labeled as dubious. These include GRBs with known properties and GRBs for which the traditional classification methods struggle in correctly identifying their progenitor.

Still, in the \textit{head}\footnote{The two substructures of the distribution were identified as \textit{tail} and \textit{head} in \cite{Negro2025}.} of the distribution, a small number (5 events) of GRBs is classified as dubious, and a significant number of GRBs fall in the 'gray area', that is the region of the correlation distribution we decided to be agnostic regarding the trustworthiness of the dimensionality reduction. This suggests that the disposition of these events in the 2D projection may not faithfully represents the structure of the original 30D distribution. Although this limitation does not compromise the main results presented in \cite{Negro2025}, it highlights that further refinements to the algorithm could improve the reliability of the 2D representation, particularly in the \textit{head} region. Work in this direction is currently underway, with the goal of ensuring that the reduced-dimensionality distribution faithfully captures the intrinsic structure of the data.

The Python script used to generate the plot shown in Fig.~\ref{figure}, along with functions that allow the algorithm to be applied to arbitrary datasets, is available at ~\url{https://github.com/nicolocibrario/embedding_reliability}.

\bibliography{rnaas}{}
\bibliographystyle{aasjournal}
\end{document}